\documentclass{ws-mpla}
\begin{document}

\newcommand{\lsim}{\mbox{\raisebox{-.9ex}{~$\stackrel{\mbox{$<$}}{\sim}$~}}}
\newcommand{\gsim}{\mbox{\raisebox{-.9ex}{~$\stackrel{\mbox{$>$}}{\sim}$~}}}

\markboth{Yeinzon Rodr\'{\i}guez}
{Low scale inflation and the immediate heavy curvaton decay}

\catchline{}{}{}{}{}

\title{LOW SCALE INFLATION AND THE IMMEDIATE HEAVY CURVATON DECAY}

\author{YEINZON RODR\'IGUEZ}
\address{Department of Physics, Lancaster University, \\
Lancaster LA1 4YB, UK \\
and \\
Centro de Investigaciones, Universidad Antonio Nari\~no, Cll 58A \# 37-94, \\
Bogot\'a D.C., Colombia \\
y.rodriguezgarcia@lancaster.ac.uk}

\maketitle

\pub{Received 2 May 2005}{Revised 14 June 2005}

\begin{abstract}
The end of a thermal inflation era, driven by the rolling of a flaton field coupled to the curvaton, cause a huge increment in the curvaton mass and decay rate while the curvaton is still frozen. It is shown that, if this increment is enough for the curvaton to immediately decay, low scale inflation with Hubble parameter $H_\ast \sim 10^3$ GeV is achieved for more natural values of the flaton-curvaton coupling constant $\lambda$ ($10^{-10} \lsim \lambda \lsim 10^{-4}$) and the curvaton bare mass $m_\sigma$ ($m_\sigma \lsim 1$ GeV).

\end{abstract}
\ccode{PACS Nos.: 98.80.Cq}

\section{Introduction}

Low scale inflation {\it is} desirable in order to identify the inflaton field with one of the MSSM flat directions \cite{kasuya} or with one of the fields appearing in the SUSY breaking sector, giving the inflaton a much deeper particle physics root. In contrast low scale inflation {\it is not} desirable because it makes very difficult the generation of the adiabatic perturbations by the inflaton, leading to multiple fine-tuning and model-building problems, unless the curvaton mechanism is invoked \cite{curv1,lyth03c,curv3,curv4,curv5,curv6}. With the aim of generating the curvature perturbation that gives origin to the large-scale structure in the observable universe, the curvaton mechanism has appeared as a nice and plausible option and a lot of research has been devoted to its study. Making the curvaton mechanism viable in a low energy inflationary framework would be the ideal situation but, unfortunately, the simplest curvaton model has shown to be incompatible with low enough values for the Hubble parameter during inflation \cite{lyth04}. Some general proposals to make the curvaton paradigm accommodate low scale inflation have recently appeared and specific models have also been studied \cite{matsuda03,postma04,yeinzon}. In a recent proposal \cite{yeinzon}, a thermal inflation epoch was attached to the general curvaton mechanism making the curvaton field gain a huge increment in the mass at the end of the thermal inflationary period, triggering this way a period of curvaton oscillations, and lowering the main inflationary scale to satisfactory levels. However, the parameters of the model required for this effect to take place showed to be extremely small to affect the reliability of the model. The purpose of this letter is to study the same mechanism but in the case where the increment in the mass is so huge that the decay rate becomes bigger than the Hubble parameter and the curvaton decays immediately. The results are very positive, offering a more natural parameter space.

\section{Thermal Inflation Model and the Curvaton Field}

Thermal inflation was introduced as a very nice mechanism to get rid of some unwanted relics that the main inflationary epoch is not able to dilute, without affecting the density perturbations generated during ordinary inflation. As its name suggests, thermal inflation relies on the finite-temperature effects on the ``flaton" scalar potential. A flaton field $\chi$ could be defined as a field with mass 
$m_\chi$ and vacuum expectation value $M \gg m_\chi$ 
\cite{lyth95,lyth96}; and 
the possible candidates within particle physics would be 
one of the many expected gauge singlets in string theory \cite{polchinski}, the GUT Higgs (which is a scalar field charged under the GUT gauge symmetry but neutral under the Standard Model one) with \hbox{$m_\chi \sim 10^3$ GeV} and $M \sim 10^{16}$ GeV \cite{lyth95}\footnote{Note, though, that in some GUT models there are additional Higgs fields with much smaller vevs \cite{guts1,guts2,guts3,guts4}.}, or the Peccei-Quinn field with (presumably) $m_\chi \sim 10^{-5}$ eV and $M \sim 10^{11}$ GeV \cite{lyth96}\footnote{In this paper we are going to focus in soft-SUSY masses. In particular, we choose a value for the gravitino mass \hbox{$m_{3/2} \sim 10^3$ GeV} which comes from gravity-mediated SUSY breaking. This means that the results of this paper do not apply for the Peccei-Quinn field as our flaton field.}.
After the period of reheating following the main inflationary epoch, the thermal background modifies the flaton potential $V$ trapping the flaton field at the origin and preventing it to roll-down towards $M$ \cite{lazarides86,barreiro96}. At this stage the total energy density $\rho_{\rm total}$ and pressure $P_{\rm total}$ are
\begin{eqnarray}
\rho_{\rm total} &=& V + \rho_T \,, \nonumber \\
P_{\rm total} &=& -V + \frac13 \rho_T \,, 
\end{eqnarray}
making the condition for thermal inflation, $\rho_{\rm total} + 3P_{\rm total} < 0$, valid when the thermal energy density $\rho_T$ falls below the height of the potential $V_0$,
which corresponds to a temperature of roughly $V_0^{1/4}$. Thermal inflation ends when the finite temperature becomes
ineffective at a temperature of order $m_\chi$, so the number of e-folds this inflationary period lasts is

\begin{equation}
N = \ln\left(\frac{a_{\rm{end}}}{a_{\rm{start}}}\right) 
=\ln \left(\frac{T_{\rm{start}}}{T_{\rm{end}}}\right)
\sim \ln\left(\frac{V_0^{1/4}}{m_\chi}\right) 
\sim\frac 12\ln\left(\frac{M}{m_\chi}\right) \sim 10 \,.
\end{equation}
Here we have used the fact that, in a flaton potential of the form
\begin{equation}
V = V_0 - (m_\chi^2 - gT^2) |\chi|^2 + \sum_{n=1}^{\infty} \lambda_n
m_P^{-2n} |\chi|^{2n+4} \,,
\end{equation}
where the $n$th term dominates:
\begin{eqnarray}
\tilde{m}_\chi^2 &=& 2 (n+1) m_\chi^2 \,, \\
M^{2n+2} m_P^{-2n} &=& [2(n+1)(n+2) \lambda_n]^{-1} \tilde{m}_\chi^2 \,, \\
V_0 &=& [2(n+2)]^{-1} \tilde{m}_\chi^2 M^2 \,,
\end{eqnarray}
with $m_P$ being the reduced Planck mass.
Note that the $gT^2$ contribution to the effective mass of the flaton field stands for the effect of the thermal background, which changes the slope of the potential in the $\chi$ direction and traps the flaton field at the origin of the potential \cite{lazarides86,barreiro96}.
It is also worthwhile to mention that the potential is stabilized by non-renormalisable terms, with dimensionless couplings $\lambda_n \sim 1$ to make the theory valid up to the Planck scale.  Otherwise, the vacuum
expectation value $M$ would not be much bigger than $\tilde{m}_\chi$, spoiling the suppression of the unwanted relics.

Before embedding the thermal inflation epoch and the curvaton mechanism into a single model, we want to clarify some issues about the nature of the interactions that produce the thermal background.
If the flaton is a GUT Higgs, it is coupled with those fields charged under the GUT gauge symmetry, in particular with those the inflaton field decays into. That collection of particles makes the thermal background, and its interaction with the flaton field produces the thermal correction. If the flaton field is a gauge singlet it still can be coupled, via Yukawa coupling terms, with some other fields, possibly in a hidden sector, that the inflaton field decays into. Again, a thermal correction is generated. The actual interactions and decay rate are not important as the main objective of this paper is to obtain some particle physics model-independent information about the possibility of reconcile low scale inflation with the curvaton mechanism, in a scenario that involves a second period of inflation (thermal inflation), without going into the details of the identification of all the relevant fields (inflaton, flaton, and curvaton) in the framework of a particle physics model (GUT theories, MSSM, etc.), which would make the results highly particle physics model-dependent. The flaton could be either a gauge singlet or the GUT Higgs.  In the former case the flaton can be coupled with some other fields that the inflaton field decays into, via Yukawa coupling terms, and the specific interactions would be known once we choose what of the many gauge singlets expected in string theory is the flaton.
In the latter case the interactions in the GUT models are already known. The specific interactions are important of course, but there are so many possibilities that the general result would be hidden behind the characteristics associated to any definite particle physics model.

Having discussed the nature of the flaton interactions, and guided by the result in Ref. \refcite{matsuda03}, we proceed to implement a second inflationary stage into the curvaton
scenario in order to lower the main inflationary energy scale \cite{yeinzon}.  If this second epoch of inflation is the thermal inflation one devised in Refs. \refcite{lyth95,lyth96,lazarides86} we would be solving not only the issue of the ordinary inflation energy scale but also the moduli problem present in the standard cosmology.

In the curvaton model supplemented by a thermal inflation epoch two fields $\chi$ and $\sigma$, which we identify as the flaton and the curvaton fields respectively, are embedded into the radiation background left by the inflaton decay. It is assumed that the curvaton field could be a gauge singlet \cite{polchinski} or a MSSM flat direction \cite{enqvistmpl,poma,kasuya2,mcdonald1,mcdonald,hamaguchi,momu,enqvist,mcdonald2,postma2,martin}\footnote{In the latter case the decay rate is suppressed enough because the field is well displaced from the origin of the potential.}, and has just a quadratic interaction with the flaton one so that it is frozen at some value $\sigma_\ast$ until the time when the flaton field is released from the origin and rolls down toward the minimum of the potential. This in turn signals the end of the thermal inflation era and the beginning of the oscillations of the curvaton field around the minimum of its quadratic potential \cite{yeinzon,CD}. The flaton field, in addition to the non renormalisable terms with $\lambda_n \sim 1$ that stabilize the potential and make its slope in the $\chi$ direction be very flat, presents a quadratic interaction with the curvaton field. The complete expression for the potential is
\begin{equation}
V(\chi, \sigma) = V_0 - (m_\chi^2 - g T^2) |\chi|^2 + m_\sigma^2 |\sigma|^2
+ \lambda |\chi|^2 |\sigma|^2 + \sum_{n=1}^\infty \lambda_n m_P^{-2n} |\chi|^{2n+4} \,,
\end{equation}
where $m_\chi \sim 10^3 \ {\rm GeV}$ due to the soft SUSY contributions in a gravity mediated SUSY breaking scheme, and $m_\sigma$ is likely to be in the range $10^{-12} \ {\rm GeV} - 10^{-1}$ GeV according to the thermal inflation model discussed in \hbox{Ref. \refcite{yeinzon}} where the curvaton field has some time to oscillate before decaying. Under these circumstances the condition for an inflationary period, $\rho_{\rm total} + 3P_{\rm total} < 0$, is satisfied when the thermal energy density $\rho_T$ falls below $V_0$. Of course, this period of thermal inflation ends when the effect of the thermal background becomes unimportant, at a temperature $T \sim m_\chi$, liberating the flaton field to roll down towards the minimum of the potential and letting it acquire a large vacuum expectation value $M$ given by:
\begin{equation}
M \simeq \frac{V_0^{1/2}}{m_\chi} \,.
\end{equation}

\begin{figure}[t]
\centerline{\psfig{file=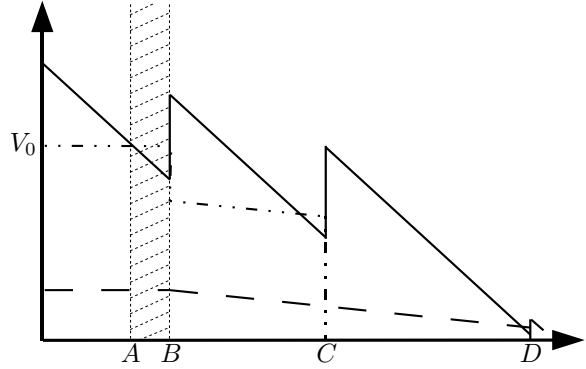,width=15cm,height=20cm}
\put(-320,470){$V_0$}
\put(-278,390){$A$}
\put(-263,390){$B$}
\put(-204,390){$C$}
\put(-127,390){$D$}}
\vspace*{-13.5cm}
\caption{Evolution of the energy densities in the thermal inflation model where the curvaton field $\sigma$ has some to oscillate before decaying (see Ref. 11). The continuos line corresponds to the radiation energy density $\rho_T$, the dashed dotted line corresponds to the flaton energy density $\rho_\chi$, and the dashed line corresponds to the curvaton energy density $\rho_\sigma$. The horizontal axis represents the expansion parameter $a$. From the left to $A$ radiation dominates the energy density, although it decreases following $\rho_T \propto a^{-4}$. At this stage the flaton and curvaton fields $\chi$ and $\sigma$ are frozen at $\chi = 0$ and $\sigma = \sigma_\ast$ making their energy densities constants. When $\rho_T$ reaches $V_0$ at $A$, thermal inflation begins. The thermal inflation period lasts until $B$ when the temperature $T$ becomes of the order of the flaton mass $m_\chi$. Thermal inflation stage is portraid by the dashed region. After thermal inflation ends, the parametric resonance process transforms a substantial fraction of $\rho_\chi$ into $\rho_T$ (see Refs. 33,34,35). The flaton field is liberated by this time and begins oscillating around the minimum of its potential, behaving then as a matter fluid with $\rho_\chi \propto a^{-3}$. The curvaton field increments suddenly its mass $m_\sigma$ at $B$ as a result of the oscillations of $\chi$ around the vacuum expectation value $M$. The increment is enough for the effective curvaton mass $\tilde{m}_\sigma$ to overtake $H_{\rm pt}$ (the Hubble parameter at $B$) so that $\sigma$ gets unfrozen and starts oscillating around $\sigma = 0$. The curvaton field behaves then as a matter fluid so that $\rho_\sigma \propto a^{-3}$. By the time $C$, $\chi$ already dominates the energy density before decaying into radiation. The curvaton field continues to oscillate until $D$ when it decays into radiation after having come to dominate (though not necessarily) the total energy density. The curvature perturbation is transfered to the radiation at this moment due to the decay of $\sigma$. This figure pretends to be just a schematic view of the different stages in the thermal inflation model discussed in Ref. 11, so that the figure is not to scale (for example, $\rho_\sigma$ is actually negligible compared with $\rho_\chi$ and $\rho_T$ on the vertical axis). \label{standard}}
\end{figure}

The thermal inflation model has been investigated before and found to be a very efficient mechanism to dilute the abundance of some unwanted relics, like the moduli fields, that the main inflationary epoch is not able to get rid of \cite{lyth95,lyth96}. We will constrain the available parameter space for $\lambda$ and $m_\sigma$ so that enough dilution of the moduli abundance is obtained.
In Ref. \refcite{yeinzon} this was done for the case in which the flaton-curvaton coupling term gives a huge contribution to the mass of the curvaton when the flaton field is released and gets its vacuum expectation value $M$. In that case the effective curvaton mass $\tilde{m}_\sigma$ may become bigger than the Hubble parameter giving birth to a period of curvaton oscillations and making the scale of the main inflationary period low enough ($H_\ast \sim m_{3/2} \sim 10^3 \ {\rm GeV}$, being $H_\ast$ the Hubble parameter when the cosmological scales leave the horizon) to think about the inflaton as a field associated to the SUSY breaking sector \cite{lyth04,matsuda03,postma04}. The evolution of the energy densities associated to the different fluids in this case are sketched in Fig. \ref{standard}.

\begin{figure}[t]
\centerline{\psfig{file=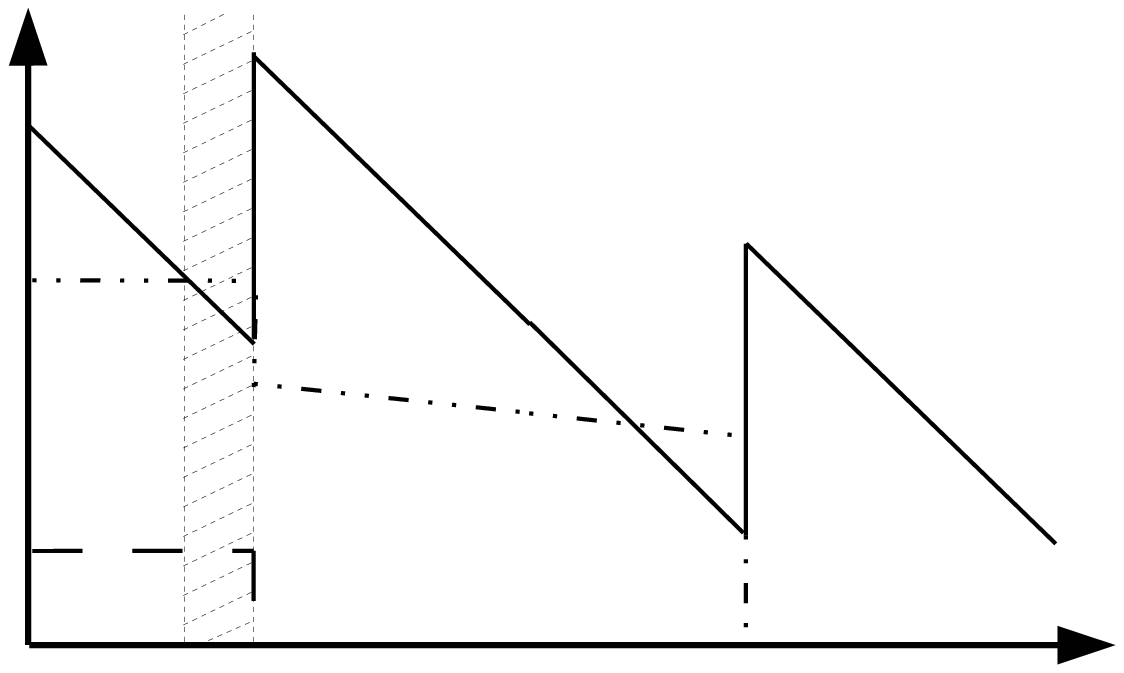,width=15cm,height=20cm}
\put(-345,470){$V_0$}
\put(-303,390){$A$}
\put(-289,390){$B$}
\put(-183,390){$C$}}
\vspace*{-13.5cm}
\caption{Evolution of the energy densities in the thermal inflation model where the curvaton field $\sigma$ decays immediatly at the end of thermal inflation. The continuos line corresponds to the radiation energy density $\rho_T$, the dashed dotted line corresponds to the flaton energy density $\rho_\chi$, and the dashed line corresponds to the curvaton energy density $\rho_\sigma$. The horizontal axis represents the expansion parameter $a$. From the left to $A$ radiation dominates the energy density, although it decreases following $\rho_T \propto a^{-4}$. At this stage the flaton and curvaton fields $\chi$ and $\sigma$ are frozen at $\chi = 0$ and $\sigma = \sigma_\ast$ making their energy densities constants. When $\rho_T$ reaches $V_0$ at $A$, thermal inflation begins. The thermal inflation period lasts until $B$ when the temperature $T$ becomes of the order of the flaton mass $m_\chi$. Thermal inflation is portraid by the dashed region. After thermal inflation ends, the parametric resonance process transforms a substantial fraction of $\rho_\chi$ into $\rho_T$ (see Refs. 33,34,35). The flaton field is liberated by this time and begins oscillating around the minimum of its potential, behaving then as a matter fluid with $\rho_\chi \propto a^{-3}$. The curvaton field increments suddenly its mass $m_\sigma$ at $B$ as a result of the oscillations of $\chi$ around the vacuum expectation value $M$. The increment is enough for the decay rate $\Gamma_\sigma$ to overtake $H_{\rm pt}$ (the Hubble parameter at $B$) so that $\sigma$ decays immediatly. The curvaton energy density is transfered then completely to $\rho_T$ as it is the curvature perturbation too. By the time $C$, $\chi$ already dominates the energy density before decaying into radiation. This figure pretends to be just a schematic view of the different stages in the thermal inflation model discussed in this paper, so that the figure is not to scale (for example, $\rho_\sigma$ is actually negligible compared with $\rho_\chi$ and $\rho_T$ on the vertical axis). \label{modified}}
\end{figure}

The purpose of this letter is to analyse the scenario where there are no oscillations of the curvaton field. As it was pointed out in Ref. \refcite{postma04} the lower bound on the curvaton decay rate
\begin{equation}
\Gamma_\sigma \geq \frac{\tilde{m}_\sigma^3}{m_P^2} \,,
\end{equation}
coming from the requirement that the curvaton interactions must be at least of gravitational strength,
is also increased when the flaton field acquires its vacuum expectation value. Thus, if this increment is big enough for the curvaton decay rate to be bigger than the Hubble parameter, the curvaton field may decay immediately rather than oscillating for some time. Low scale inflation in this case is also possible to be attained \cite{postma04,yeinzon}, but the lower bound on $H_\ast$ changes with respect to the case when the curvaton oscillatory process is triggered. The evolution of the energy densities associated to the different fluids in this case are sketched in Fig. \ref{modified}.

\section{The Bounds on the Scale of Inflation} \label{mbsi}

In this section we present four bounds on the scale of inflation, in terms
of two parameters which encode possible modifications of the simplest 
curvaton scenario. These bounds have been presented at least implicitly in
earlier works \cite{lyth04,matsuda03,postma04}, but only Ref. \refcite{yeinzon}
presents them in the unified notation that we employ. The advantage of this notation is that it
will allow us to compare the 
bounds in various situations, establishing with ease which is  the most 
crucial. The two parameters are
\begin{itemize}
\item
The ratio $f\equiv H_{\rm pt}/\tilde m_\sigma$, where $H_{\rm pt}$ is the 
Hubble parameter at the end of thermal inflation and $\tilde m_\sigma$
is the effective curvaton mass after the end of thermal inflation.
\item
The ratio $\delta\equiv\sqrt{H_{\rm pt}/H_*}$ where $H_*$ is the Hubble 
parameter during inflation.
\end{itemize}

\subsection{Curvaton physics considerations}

The  observed value of the nearly scale invariant spectrum of curvature 
perturbations, parameterised by the amplitude $A_\zeta$, is 
$A_\zeta \approx 5\times 10^{-5}$ \cite{observation}.
In the curvaton scenario $\zeta$ is given by \cite{curv1,lyth03c}
\begin{equation}
\zeta \approx \Omega_{\rm dec}\zeta_\sigma \,,
\label{zeta}
\end{equation}
where \mbox{$\Omega_{\rm dec}\leq 1$} is the density fraction of the curvaton energy
density $\rho_\sigma$ over the total energy density of the Universe $\rho_{\rm total}$ at the time
of the decay of the curvaton:
\begin{equation}
\Omega_{\rm dec}\equiv\left(\frac{\rho_\sigma}{\rho_{\rm total}}\right)_{\rm dec} \,,
\label{r}
\end{equation}
and $\zeta_\sigma$ is the curvature perturbation of the curvaton field
$\sigma$, which is in the flat slice \cite{CD}
\begin{equation}
\zeta_\sigma\sim\left(\frac{\delta\sigma}{\sigma}\right)_{\rm dec}\approx
\left(\frac{\delta\sigma}{\sigma}\right)_{\rm pt} \,,
\label{zs1}
\end{equation}
where `pt' denotes the time at the end of the thermal inflation period and `dec'
denotes the time of curvaton decay.

In our thermal inflation model
\begin{equation}
\left(\frac{\delta\sigma}{\sigma}\right)_*\simeq
\left(\frac{\delta\sigma}{\sigma}\right)_{\rm pt} \,,
\label{ds/s}
\end{equation}
where `*' denotes the epoch when the cosmological scales exit the horizon 
during inflation. The above typically holds true because the curvaton (being a 
light field) is frozen during and after inflation until the end of the thermal
inflation period (see Figs. \ref{standard} and \ref{modified}).

Now, for the perturbation of the curvaton we have the following value for the
amplitude $A_{\delta \sigma_\ast}$ of the spectrum of perturbations \cite{bunch}
\begin{equation}
A_{\delta\sigma_*}=\frac{H_*}{2\pi}\,,
\label{dsH}
\end{equation}
which from Eqs.~(\ref{zeta}), (\ref{zs1}), and (\ref{ds/s}) leads to
\begin{equation}
\sigma_\ast \sim \Omega_{\rm dec}\,\frac{\delta\sigma_\ast}{\zeta} = \Omega_{\rm dec} \frac{A_{\delta \sigma_\ast}}{A_\zeta} \,.
\end{equation}
Using Eq.~(\ref{dsH}), we can recast the above as
\begin{equation}
\sigma_\ast \sim\frac{H_*\Omega_{\rm dec}}{\pi A_\zeta} \,.
\label{sosc}
\end{equation}

\subsection{The main bound on the scale of inflation}

For the density fraction at the end of thermal inflation we have:
\begin{equation}
\left(\frac{\rho_\sigma}{\rho_{\rm total}}\right)_{\rm pt}\sim f^{-2}
\left(\frac{\sigma_\ast}{m_P}\right)^2 \,,
\label{rhofracosc}
\end{equation}
where 
\begin{equation}
f\equiv\frac{H_{\rm pt}}{\tilde m_\sigma} \,,
\label{fff}
\end{equation}
and
we used that 
\mbox{$(\rho_\sigma)_{\rm pt}\simeq
\frac{1}{2}\tilde m_\sigma^2\sigma_\ast^2$} 
and \mbox{$(\rho_{\rm total})_{\rm pt}=3H_{\rm pt}^2m_P^2$}. Here, $\tilde m_\sigma$
denotes the effective mass of the curvaton {\em after} the end of thermal inflation.
In the basic setup of the curvaton hypothesis this effective mass
 is the bare mass $m_\sigma$. If this is the case then 
\mbox{$\tilde m_\sigma=m_\sigma\simeq H_{\rm pt}$} (i.e. \mbox{$f\simeq 1$}).
However, in the heavy curvaton scenario, the mass of the curvaton
is supposed to be suddenly incremented at some time after the end of the 
inflationary epoch due to a coupling of the form 
$\lambda \chi^2 \sigma^2$ with a field $\chi$ which acquires a large vacuum
 expectation value at some time after the end of inflation \cite{lyth04,yeinzon}. In this case
\mbox{$\tilde{m}_\sigma^2=m_\sigma^2+\lambda\langle\chi\rangle^2\approx 
\lambda\langle\chi\rangle^2\gg H_{\rm pt}^2$} (\hbox{i.e. \mbox{$f\ll 1$}}).

Now, we need to consider separately the cases when the curvaton decays before
it dominates the Universe (\mbox{$\Omega_{\rm dec}\ll 1$}) or after it does
so (\mbox{$\Omega_{\rm dec}\sim 1$})\footnote{The former case includes both the possibility that the curvaton oscillates for some time before decaying and the possibility that the curvaton decays immediatly. The latter case is valid only if the curvaton oscillates before decaying.}. Note, that the WMAP constraints 
on non-gaussianity in the CMB impose a lower bound on $\Omega_{\rm dec}$, which allows the
range \cite{lyth03c,komatsu03}
\begin{equation}
0.01 \leq \Omega_{\rm{dec}} \leq 1 \,. 
\label{WMAPr}
\end{equation} 
Because of the above bound we require that the density ratio $\rho_\sigma/\rho_{\rm total}$
grows substantially after the end of inflation. Typically, in
the curvaton scenario this does indeed take place after the curvaton begins
oscillating, but only if the curvaton oscillates in a quadratic potential 
during the radiation era. As it was shown in Ref.~\refcite{CD}, if the curvaton 
oscillates in a quartic or even higher order potential, its density ratio does 
not increase with time (it may well decrease instead) and satisfying the bound 
in Eq.~(\ref{WMAPr}) is very hard. Due to this fact, in the following,
we assume that
the period of oscillations 
occurs in the radiation era with a quadratic potential. Hence, we consider that
\mbox{$H_{\rm pt}\leq\Gamma_{\rm inf}$}, being $\Gamma_{\rm inf}$ the inflaton
decay rate.

Suppose, at first, that the curvaton decays before dominating the energy density of
the Universe so that \mbox{$\Omega_{\rm dec}\ll 1$}.
Assuming that the curvaton oscillates in a quadratic potential, during the 
radiation epoch, its density fraction grows as 
\mbox{$\rho_\sigma/\rho_{\rm total}\propto H^{-1/2}$}. 
Therefore, at curvaton decay we have
%
%
\begin{equation}
\Omega_{\rm{dec}}\sim
\frac{\tilde m_\sigma^2\sigma_\ast^2}{T_{\rm{dec}}
H_\ast^{3/2}m_P^{3/2}} \,, 
\label{Tdec}
\end{equation}
where we used Eq.~(\ref{rhofracosc}) and also that 
\mbox{$(\rho_{\rm total})_{\rm dec}\sim T_{\rm dec}^4$}.
Using Eq.~(\ref{sosc}) the above can be recast as
%
%
\begin{equation}
H_*\sim\pi A_\zeta f
\frac{m_P}{\sqrt{\Omega_{\rm dec}}}
\left(\frac{H_{\rm dec}}{H_{\rm pt}}\right)^{1/4} \,,
\label{H*1}
\end{equation}
where we used that \mbox{$T_{\rm dec}^2\sim H_{\rm dec}m_P$}.

Now, suppose that the curvaton decays after it dominates the Universe so that
\mbox{$\Omega_{\rm dec}\sim 1$}. Since 
\mbox{$(\rho_\sigma/\rho_{\rm total})_{\rm dom}\simeq 1$}
by definition, using again that, during the radiation epoch, 
\mbox{$\rho_\sigma/\rho_{\rm total}\propto H^{-1/2}$} and in view of 
Eq.~(\ref{rhofracosc}), we obtain
%
%
\begin{equation}
H_{\rm dom}\sim H_{\rm pt} f^{-4}
\left(\frac{\sigma_\ast}{m_P}\right)^4 \,,
\label{Hdom}
\end{equation}
where `dom' denotes the time of curvaton domination\footnote{Here we define $H_{\rm dom}$ by $H_{\rm dom} = H_{\rm eq}$, where $H_{\rm eq}$ is the Hubble parameter at the time when the curvaton energy density $\rho_\sigma$ makes equal to the radiation energy \hbox{density $\rho_T$}.}. Employing again 
Eq.~(\ref{sosc}), the above can be written as
%
%
\begin{equation}
H_*\sim\pi A_\zeta
f m_P
\left(\frac{H_{\rm dom}}{H_{\rm pt}}\right)^{1/4} \,.
\label{H*2}
\end{equation}

Combining Eqs.~(\ref{H*1}) and (\ref{H*2}) we find that, in all cases
%
%
\begin{equation}
H_* \sim \pi A_\zeta f\frac{m_P}{\sqrt{\Omega_{\rm dec}}}
\left(\frac{\max\{H_{\rm dom}, H_{\rm dec}\}}{H_{\rm pt}}\right)^{1/4} \,.
\label{H*0}
\end{equation}
This can be rewritten as 
%
%
\begin{equation}
H_* \sim \Omega_{\rm dec}^{-2/5}
\left(\frac{H_*}{H_{\rm pt}}\right)^{1/5}
\left(\frac{\max\{H_{\rm dom}, H_{\rm dec}\}}{H_{\rm BBN}}\right)^{1/5}
(\pi A_\zeta f)^{4/5}(T_{\rm BBN}^2m_P^3)^{1/5} \,,
\label{H*}
\end{equation}
where `BBN' denotes the epoch of Big Bang Nucleosynthesis (BBN)
(\mbox{$T_{\rm BBN}\sim 1$ MeV}).
Now, according to Eq.~(\ref{WMAPr}) we have \mbox{$\Omega_{\rm dec}\leq 1$}.
Also, we require that the curvaton decays before BBN, i.e. 
\mbox{$H_{\rm dec}>H_{\rm BBN}$}.
Hence, the above provides the following bound
\begin{equation}
\mbox{\framebox{%
\begin{tabular}{c}\\
$H_*>(\pi A_\zeta f)^{4/5}(T_{\rm BBN}^2m_P^3)^{1/5}\sim
f^{4/5}\times 10^7 \ {\rm GeV}$.\\
\\
\end{tabular}}}
\label{H*bound}
\end{equation}
In the standard setup of the curvaton scenario \mbox{$f=1$} and
the above bounds do not allow inflation at low energy scales to take place
\cite{lyth04}. However, we see that if $f$ is much
smaller than unity the lower bound on the inflationary scale can be 
substantially relaxed and low scale inflation can be accommodated. Still, 
though, there are more bounds to be considered.

\subsection{Other bounds related to curvaton decay}

Firstly, let us consider the bound coming from the fact that the decay rate
of the curvaton field cannot be arbitrarily small. Indeed, in view of 
the fact that the curvaton interactions are at least of gravitational strength,
we find the following decay rate for the curvaton
\begin{equation}
\Gamma_\sigma \approx 
\gamma_\sigma 
\frac{\tilde{m}_\sigma^3}{m_P^2}\leq\tilde m_\sigma \,, 
\label{decay_rate}
\end{equation}
where \mbox{$\gamma_\sigma \gsim 1$}. 

Suppose, at first, that the curvaton decays after the onset of its 
oscillations, as in the basic setup of the curvaton scenario (see Fig. \ref{standard}). In this case, 
\mbox{$\Gamma_\sigma\leq H_{\rm pt}$} and \mbox{$H_{\rm dec}=\Gamma_\sigma$}.
Using the fact that 
\mbox{max$\{H_{\rm dom}, \Gamma_\sigma\}\geq\Gamma_\sigma$},
Eq.~(\ref{decay_rate}) suggests
\begin{equation}
\frac{\max\{H_{\rm dom}, H_{\rm dec}\}}{H_{\rm pt}}\geq 
\gamma_\sigma 
f^{-1}
\left(\frac{\tilde m_\sigma}{m_P}\right)^2 \,.
\end{equation}
Including the above into Eq.~(\ref{H*0}) the latter becomes
%
%
\begin{equation}
H_* \geq 
\sqrt{\gamma_\sigma}(\pi A_\zeta)^2\sqrt{f}\,
\frac{m_P}{\Omega_{\rm dec}}\left(\frac{H_{\rm pt}}{H_*}\right) \,,
\label{Hbound0}
\end{equation}
which results in the bound
\begin{equation}
\mbox{\framebox{%
\begin{tabular}{c}\\
$H_*\geq 
(\pi A_\zeta)^2\sqrt{f}\;\delta^2\,m_P
\sim
\sqrt f\;\delta^2
\times 10^{11} \ {\rm GeV}$, \\
\\
\end{tabular}}}
\label{Hbound}
\end{equation}
where we have defined 
\begin{equation}
\delta\equiv
\sqrt{\frac{H_{\rm{pt}}}{H_\ast}}\leq 1 \,.
\label{dratio}
\end{equation}

Now, provided we demand that the curvaton field does not itself result in a
period of inflation, we see that the curvaton cannot dominate the Universe 
before the end of the thermal inflationary period. This results into the constraint
\begin{equation}
\left(\frac{\rho_\sigma}{\rho_{\rm total}}\right)_{\rm pt} \leq 1
\Leftrightarrow
\tilde m_\sigma\leq\pi A_\zeta\,\delta^2\frac{m_P}{\Omega_{\rm dec}}
\Leftrightarrow
f\geq\frac{\Omega_{\rm dec}H_*}{(\pi A_\zeta)m_P} \,,
\label{cons_m_n}
\end{equation}
where we used Eqs.~(\ref{sosc}), (\ref{rhofracosc}), (\ref{fff}) and 
(\ref{dratio}). Inserting the above into Eq.~(\ref{Hbound0}) we obtain
%
%
\begin{equation}
H_*\geq
\gamma_\sigma(\pi A_\zeta)^3\delta^4\,
\frac{m_P}{\Omega_{\rm dec}} \,,
\label{Hbound1}
\end{equation}
which results in the bound
\begin{equation}
\mbox{\framebox{%
\begin{tabular}{c}\\
$H_*\geq (\pi A_\zeta)^3\delta^4m_P\sim
\delta^4\times 10^7 \ {\rm GeV}$. \\
\\
\end{tabular}}}
\label{Hbound-0}
\end{equation}
A similar bound is reached with the use of the upper bound on 
$\tilde{m}_\sigma$
\begin{equation}
\tilde m_\sigma\leq\gamma_\sigma^{-1/3}(H_{\rm pt}m_P^2)^{1/3} \,,
\label{msbound}
\end{equation}
which comes from $\Gamma_\sigma\leq H_{\rm pt}$ and the 
Eq.~(\ref{decay_rate}), 
instead of the bound in Eq. (\ref{cons_m_n}). Inserting the above into 
Eq.~(\ref{Hbound0}) one finds [cf. Eq.~(\ref{Hbound1})]
%
%
\begin{equation}
H_*\geq
\gamma_\sigma(\pi A_\zeta)^3\delta^4\,
\frac{m_P}{\Omega_{\rm dec}^{3/2}} \,,
\label{Hbound2}
\end{equation}
which, again, results in the bound in Eq.~(\ref{Hbound-0}), as it was 
suggested in Ref.~\refcite{postma04}.

If \mbox{$\delta\rightarrow 1$}, the bounds in 
Eq.~(\ref{Hbound-0})
are not possible to be relaxed below the standard case discussed in 
Ref.~\refcite{lyth04} despite the fact that we may have \mbox{$f\ll 1$} in 
Eq.~(\ref{H*bound}). 
Therefore, in the heavy curvaton 
scenario we require \mbox{$\delta\ll 1$}, i.e. {\em the onset of the curvaton
oscillations has to occur much later than the end of inflation} so that 
\mbox{$H_*\gg H_{\rm pt}\geq\Gamma_\sigma$} \cite{matsuda03}. In this case, as can be seen in
Eq.~(\ref{Hbound-0}), 
it is easy to lower the bound
on the inflationary scale even for a not-so-small value of $\delta$.
This is a very nice feature of this scenario.

As it was pointed out in Ref.~\refcite{postma04}, the sudden increment in the 
curvaton mass might lead to a growth in the curvaton decay rate enough for 
\mbox{$\Gamma_\sigma>H_{\rm pt}$}. This would force the curvaton to decay 
immediately (see Fig. \ref{modified}) and we can write 
\mbox{$H_{\rm pt}\sim H_{\rm dec}$}.
Obviously, in this case we cannot have \mbox{$H_{\rm dec}<H_{\rm dom}$} and
there is no period when \mbox{$\rho_\sigma/\rho_{\rm total}\propto H^{-1/2}$}. This 
means that \mbox{$(\rho_\sigma/\rho_{\rm total})_{\rm pt}\sim\Omega_{\rm dec}$}. Using
Eqs.~(\ref{sosc}) and (\ref{rhofracosc}) it is easy to find
\begin{equation}
H_*\sim\pi A_\zeta f
\frac{m_P}{\sqrt{\Omega_{\rm dec}}} \,,
\label{H*00}
\end{equation}
which results in the following bound
\begin{equation}
\mbox{\framebox{%
\begin{tabular}{c}\\
$H_*\geq\pi A_\zeta f\,m_P\sim f\times 10^{14} 
\ {\rm GeV}$. \\
\\
\end{tabular}}}
\label{Hbound-00}
\end{equation}
It is evident that the above bound may challenge the WMAP constraint for the
curvaton scenario \cite{liber} leading to excessive curvature 
perturbations from the inflaton field if $f$ is not much 
smaller than unity. 

%

The bounds in Eqs. (\ref{H*bound}), (\ref{Hbound}), and  (\ref{Hbound-0}) 
 provide the basis for the thermal inflation scenario studied in \hbox{Ref. \refcite{yeinzon}}
 where the curvaton oscillates for some time before decaying (see Fig. \ref{standard}).
 Instead, the bound in Eq.
(\ref{Hbound-00}), together with \hbox{Eq. (\ref{H*bound})}, will be considered in the next chapter
where the curvaton decays immediatly just after the end of thermal inflation (see Fig. \ref{modified}).
As a matter of completeness we have considered all the other 
possible bounds coming from the requirements that 
\mbox{$\Gamma_\sigma < \tilde{m}_\sigma$} and 
\mbox{$H_{\rm dec} \geq H_{\rm BBN}$}. We have found that these bounds lead
to consistent and/or weaker constraints than the above four.

\section{The Immediate Heavy Curvaton Decay}

In the scenario where curvaton oscillations are allowed, corresponding to $\Gamma_\sigma < H_{\rm pt}$, the lower bound on $H_\ast$ is [\hbox{c.f. Eqs. (\ref{H*bound}), (\ref{Hbound}), and (\ref{Hbound-0})}]
\begin{equation}
H_\ast \geq {\rm max} \{ \ f^{4/5} \times 10^7 \ {\rm GeV}, \sqrt f \delta^2 \times 10^{11} \ {\rm GeV},
\delta^4 \times 10^{7} \ {\rm GeV} \ \} \,, \label{H_bound}
\end{equation}
where $H_{\rm pt}$ corresponds also to the beginning of the curvaton oscillations. In contrast, the lower bound in the scenario where the curvaton field decays immediately, corresponding to $\Gamma_\sigma > H_{\rm pt}$, is [c.f. Eqs. (\ref{H*bound}) and (\ref{Hbound-00})]
\begin{equation}
H_\ast \geq {\rm max} \{ \ f^{4/5} \times 10^7 \ {\rm GeV}, f \times 10^{14} \ {\rm GeV} \ \} \,. \label{H_bound_n}
\end{equation}

The lower bound in Eq. (\ref{H_bound}), for $H_\ast \sim 10^3$ GeV, was shown \cite{yeinzon} to be satisfied for very small values for the flaton-curvaton coupling constant, $\lambda \sim 10^{-22} - 10^{-10}$, and very small values for the bare mass of the curvaton field, $m_\sigma \sim 10^{-12} \ {\rm GeV} - 10^{-1} \ {\rm GeV}$, which suggests that the curvaton field could be a pseudo Nambu-Goldstone boson \cite{chun,dlnr,hofmann}. This is, in any case, a quite negative result due to the required smallness of the parameters $\lambda$ and $m_\sigma$. However, when taking into account the lower bound in Eq. (\ref{H_bound_n}), corresponding to the case when the decay rate $\Gamma_\sigma$ becomes bigger than $H_{\rm pt}$, things change appreciably.

\subsection{The flaton-curvaton coupling constant $\lambda$}
\begin{figure}[t]
\centerline{\psfig{file=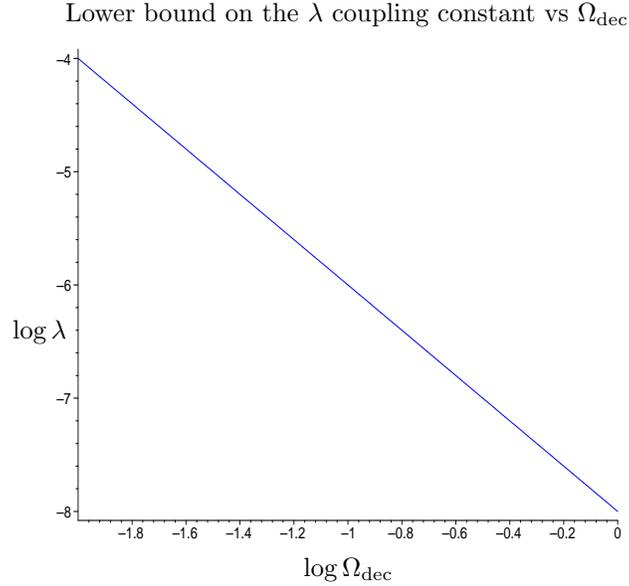,width=6.5cm,height=7.5cm,angle=-90}
\put(-210,10){Lower bound on the $\lambda$ coupling constant vs $\Omega_{\rm dec}$}
\put(-230,-110){$\log{\lambda}$}
\put(-120,-200){$\log{\Omega_{\rm dec}}$}}
\vspace*{8pt}
\caption{Lower bound on the flaton-curvaton coupling constant $\lambda$ as a logarithmic plot. A more natural value for $\lambda$ requires a higher level of non gaussianity compatible with the WMAP constraints. \label{fig}}
\end{figure}

Thermal inflation ends when the thermal energy density is no longer dominant; thus, the Hubble parameter at the end of thermal inflation is associated to the energy density coming from the curvaton and the flaton fields:
\begin{equation}
H_{\rm{pt}}^2 = \frac{\rho_T + V(\chi = 0, \sigma = \sigma_\ast)}{3 m_P^2} \sim \frac{m_\chi^2 M^2}{3 m_P^2} \,,
\end{equation}
so that
\begin{equation}
H_{\rm{pt}} \sim 10^{-16} M \,. \label{tic}
\end{equation}
Since the effective mass of the curvaton field after the end of thermal inflation, i.e., when $\bar \chi = M_\chi$ and $\bar \sigma = 0$ are the average values of the flaton and the curvaton fields, is
\begin{equation}
\tilde{m}_\sigma = (m_\sigma^2 + \lambda M^2)^{1/2} \approx \sqrt{\lambda} M \,, \label{cef}
\end{equation}
the parameter $f$ [cf. Eq. (\ref{fff})] becomes
\begin{equation}
f \equiv \frac{H_{\rm pt}}{\tilde{m}_\sigma} \sim 10^{-16} \frac{1}{\sqrt{\lambda}} \,. \label{ff}
\end{equation}
In view of  the Eqs. (\ref{H_bound_n}) and (\ref{ff}) the smallest possible value for $\lambda$, compatible with $H_\ast \sim 10^3 \ {\rm GeV}$, becomes $\lambda \sim 10^{-10}$, which is very good because already improves the results found in Ref. \refcite{yeinzon}. Moreover, the effective flaton mass during thermal inflation $\tilde{m}_\chi = (m_\chi^2 - \lambda \sigma_\ast^2)^{1/2}$ must be positive to trap the flaton field at the origin of the potential. Thus, $\lambda \sigma_\ast^2 < m_\chi^2$, and the biggest possible value for $\lambda$ becomes [\hbox{c.f. Eq. (\ref{sosc})}]
\begin{equation}
\lambda < \frac{m_\chi^2}{\sigma_\ast^2} \sim \frac{10^{-2} \ {\rm GeV}^2}{\Omega_{\rm dec}^2 H_\ast^2} \sim 10^{-4} \,, 
\end{equation}
which is already a small value but much bigger and more natural than that found in the case where curvaton oscillations are allowed \cite{yeinzon}. The lower bound on \mbox{$\lambda$ vs $\Omega_{\rm dec}$}
is depicted in Fig. \ref{fig}. Note that a small value for $\Omega_{\rm dec}$, which is restricted to be $\Omega_{\rm dec} \geq 0.01$ in order to satisfy the WMAP constraints on non gaussianity \cite{lyth03c,komatsu03}, is desirable to obtain a higher value for $\lambda$, so the biggest possible value $\lambda \sim 10^{-4}$ is at the expense of a high level of non gaussianity.

Recalling, in the scenario where curvaton oscillations are allowed the coupling constant $\lambda$ is in the range
\begin{equation}
10^{-22} \lsim \lambda \lsim 10^{-10} \,,
\end{equation}
whereas in the scenario where the curvaton decays immediately the range is
\begin{equation}
10^{-10} \lsim \lambda \lsim 10^{-4} \,.
\end{equation}

\subsection{The bare curvaton mass $m_\sigma$}
The only bound on $m_\sigma$ is given by the fact that in the heavy curvaton scenario the bare mass must be smaller than the Hubble parameter at the end of the thermal inflation era, so that the sudden increment in the mass and the decay rate leads to the immediate decay of the field avoiding in this case the oscillations. Thus,
\begin{equation}
m_\sigma < H_{\rm pt} \sim 10^{-16} M \,,
\end{equation}
so we need to worry about the possible values for $M$. In the scenario where the curvaton field decays immediately the flaton field is left immersed in a background of radiation, so it must decay before the time of nucleosynthesis in order not to disturb the abundances of the light elements. By setting $\Gamma_\chi \simeq H$ we get the temperature just after the flaton decay
\begin{equation}
T_\chi \simeq 10^{13} \ {\rm GeV}^2 \frac{1}{M} \,, \label{nucl}
\end{equation}
which must be bigger than $1$ MeV to satisfy the nucleosynthesis constraint. Therefore
\begin{equation}
M \lsim 10^{16} \ {\rm GeV} \,, \label{bound_M}
\end{equation}
leading to a lower bound on the bare curvaton mass given by $m_\sigma \lsim 1 \ {\rm GeV}$, which is again a more relaxed constraint than that found in Ref. \refcite{yeinzon} for the case of an oscillating curvaton. Recalling, in the scenario where curvaton oscillations are allowed the bare curvaton mass $m_\sigma$ is in the range
\begin{equation}
10^{-12} \ {\rm GeV} \lsim m_\sigma \lsim 10^{-1} \ {\rm GeV} \,,
\end{equation}
whereas in the scenario where the curvaton decays immediately the range is
\begin{equation}
m_\sigma \lsim 1 \ {\rm GeV} \,. \label{bound_m}
\end{equation}

Some important constraints might come from the solution to the moduli problem and could limit the reliability of the Eqs. (\ref{bound_M}) and (\ref{bound_m}). Moduli fields are flaton
fields with a vacuum expectation value $\Phi_0$ of order the Planck mass. The decay of the flaton field
increments the entropy density $s$, so that the big-bang moduli abundance, defined as that produced before thermal inflation and given by \cite{lyth96}
\begin{equation}
\frac{n_\Phi}{s} \sim \frac{\Phi_0^2}{10 m_P^{3/2} m_\Phi^{1/2}} \,,
\end{equation}
where $m_\Phi$ is the mass of the moduli fields, gets suppressed by two factors. One is
\begin{equation}
\Delta_{PR} \sim \frac{g_\ast(T_{PR})}{g_\ast(T_C)} \frac{T_{PR}^3}{T_C^3} \,,
\end{equation} 
due to the parametric resonance process \cite{kofmanpr1,kofmanpr2,traschen} following the end of the thermal inflation era, where the $g_\ast$ are the total internal particle degrees of freedom, $T_{PR}$ is the temperature just after the period of preheating, and $T_C \sim m_\chi$ is the temperature at the end of thermal inflation; the other is
\begin{equation}
\Delta_\chi \sim \frac{4\beta V_0 / 3T_\chi}{(2 \pi^2 / 45) g_\ast(T_{PR}) T_{PR}^3} \,,
\end{equation} 
due to the flaton decay, where $T_\chi$ is the temperature just after the decay\footnote{This is assuming that the flaton has come to dominate the energy density just before decaying (see Fig. \ref{modified}).}, and $\beta$ is the fraction of the total
energy density left in the flatons by the parametric resonance process ($\beta \sim 1$). Thus, the abundance of the big-bang moduli after thermal inflation is:
\begin{eqnarray}
\frac{n_\Phi}{s} & \sim &
\frac{\Phi_0^2}{10 m_P^{3/2} m_\Phi^{1/2} \Delta_{PR} \Delta_\chi} \sim \frac{10 \Phi_0^2 T_\chi T_C^3}{\beta V_0 m_\Phi^{1/2} m_P^{3/2}} \sim \nonumber \\
& \sim & 10^6 \ {\rm GeV}^2 M^{-2} \left(\frac{\Phi_0}{m_P}\right)^2%
\left(\frac{T_\chi}{1 \hspace{1mm} {\rm MeV}}\right) \left(\frac{T_C}{m_\Phi}\right)^3 \times \nonumber \\
&& \times \left(\frac{m_\Phi}{10^3 \hspace{1mm}{\rm GeV}}\right)^{1/2}%
\left(\frac{1}{\beta}\right) \left(\frac{m_\Phi^2 M^2}{V_0}\right) \,, \label{abundance}
\end{eqnarray}
which must be suppressed enough ($n_\Phi / s \lsim 10^{-12}$) so that the nucleosynthesis constraints studied in Ref. \refcite{ellis92} are satisfied. This is easily achieved by imposing an upper bound on $M$:
\begin{equation}
M \gsim 10^9 \ {\rm GeV} \,,
\end{equation} 
which does not affect the lower bounds on $M$ and $m_\sigma$ in Eqs. (\ref{bound_M}) and (\ref{bound_m}).

We also have to take care about the abundance of the thermal inflation moduli, defined as that produced after thermal inflation:
\begin{eqnarray}
\frac{n_\Phi}{s} & \sim  &
\frac{\Phi_0^2 V_0^2 / 10 m_\Phi^3 m_P^4}{(2 \pi^2 / 45) g_\ast(T_{PR}) T_{PR}^3%
\Delta_\chi} \sim \frac{\Phi_0^2 V_0 T_\chi}{ 10 \beta m_\Phi^3 m_P^4} \sim
\nonumber \\
& \sim & 10^{-44} \hspace{1mm} {\rm GeV}^{-2} M^2 \left(\frac{\Phi_0}{m_P}\right)^2 \left(\frac{T_\chi}{1 \hspace{1mm} {\rm MeV}} \right) \times \nonumber \\
&& \times \left(\frac{1}{\beta}\right) \left(\frac{10^3 \hspace{1mm} {\rm GeV}}{m_\Phi}\right) \left(\frac{V_0}{m_\Phi^2 M^2}\right) \,. \label{supp_abun_2}
\end{eqnarray}
To suppress the thermal inflation moduli at the required level $n_\Phi / s \lsim 10^{-12}$ we require
\begin{equation}
M \lsim 10^{16} \ {\rm GeV} \,, \label{entropyb2}
\end{equation}
which is precisely the same bound as in Eq. (\ref{bound_M}).  Recalling, the allowed range of values for the vacuum expectation value of the flaton field is
\begin{equation}
10^9 \ {\rm GeV} \lsim M \lsim 10^{16} \ {\rm GeV} \,,
\end{equation}
so that the moduli problem is solved and, in the best case, $m_\sigma \sim 1$ GeV.


\section{Some Useful Remarks}
Before concluding, we want to stress some points that can help to avoid possible confusion. The parameter space compatible with low scale inflation is a feature of the specific model studied, and we cannot say it is the same for all kind of models in the basis of \hbox{Eqs. (\ref{H*bound}), (\ref{Hbound}), and (\ref{Hbound-0})} \cite{yeinzon}, which provide just some general bounds. That is why specific models have been studied (\hbox{see Refs. \refcite{postma04,yeinzon}}), even when the general bounds were already known from Refs. \refcite{lyth04,matsuda03}.
Although the claim, that the available parameter space is bigger for the immediate curvaton decay,
was given before in \hbox{Ref. \refcite{postma04}}, we again cannot say that the available parameter space is the same for all kind of models in the basis of the bounds required to have low energy scale inflation.  For example, from Eq. (\ref{tic}), $H_{\rm pt}$ depends on $M$ so there is no a direct bound on it unless we know the bound on $M$ \footnote{Notice that the bounds required to have low energy scale inflation [c.f. Eq. (\ref{H_bound_n})] depend only on the ratio \mbox{$f = H_{\rm pt}/\tilde{m}_\sigma$}, and not exclusively on $H_{\rm pt}$.}. The bound on $M$ comes in turn from the requirement that the flaton decays before nuclosynthesis [c.f. \hbox{Eqs. (\ref{nucl}) and (\ref{bound_M})}] and must be consistent with the adequate suppression of the thermal inflation moduli [\hbox{c.f. Eqs. (\ref{supp_abun_2}) and (\ref{entropyb2})}]. These are, of course, features specific only to the model we are studying, and are therefore not present in Ref. \refcite{postma04}.

Naively, one would think that the bounds on $\lambda$ and $m_\sigma$ are found from that on $H_{\rm pt}$ only through a mere change of variables. This is of course not true as the bound on $H_{\rm pt}$ is a very sensible quantity that has to satisfy the not disturbance of the nucleosynthesis process and the adequate moduli abundance suppression. It is worth mentioning that the scenario discussed in this paper differs appreciably from that studied in Ref. \refcite{yeinzon}, due to the immediate curvaton decay, so that the conditions to satisfy the nuclosynthesis and thermal inflation moduli constraints are completely different\footnote{For example, the expressions for the big-bang and thermal inflation moduli abundances in Ref. \refcite{yeinzon} [\hbox{c.f. Eqs. (4.41) and (4.43)} in that reference] are different from those in this paper [\hbox{c.f. Eqs. (\ref{abundance}) and (\ref{supp_abun_2})}].}. 

Finally, the agreement between the bounds found in Ref. \refcite{postma04} (which are supposed to be general) and those found in this paper is apparent and corresponds just to a mere coincidence.  We justify this observation by noting  that Eq. (6) in Ref. \refcite{postma04} is essentially the same as our \hbox{Eq. (\ref{H*1})}, the latter being generalized to give \hbox{Eq. (\ref{H*bound})}, except for $\Gamma_\sigma$ which in our Eq. (\ref{H*1}) appears to be $H_{\rm dec}$. The expressions in \hbox{Section \ref{mbsi}} (see Ref. \refcite{yeinzon}) were carefully derived so that the correct expression is that given there. In contrast, Eq. (6) in \hbox{Ref. \refcite{postma04}} is just valid for the standard case where the curvaton field has some time to oscillate before decaying, so we can identify $\Gamma_\sigma$ with $H_{\rm dec}$. However, for the immediate decay case, $\Gamma_\sigma > H_{\rm dec} = H_{\rm pt}$, which renders \hbox{Eq. (6)} in Ref. \refcite{postma04} invalid.
Based on the previous discussion we claim that the bound \mbox{$H_{\rm pt} < 1$ GeV}, as are those on $\lambda$ and $m_\sigma$, is presented in this paper for the first time {\em in a correct way}.

\section{Conclusions}

In this letter we have investigated the required parameter space compatible with low scale inflation, in the thermal inflation curvaton scenario where there are no oscillations of the curvaton field. We have shown that the parameter space is greatly enhanced when the increment in the curvaton decay rate is big enough for the curvaton field to decay immediately at the end of the thermal inflation era. The best case corresponds to a flaton-curvaton coupling constant $\lambda \sim 10^{-4}$ and a bare curvaton mass $m_\sigma \sim 1 \ {\rm GeV}$, which are much bigger and more natural than the ranges $10^{-22} \lsim \lambda \lsim 10^{-10}$ and $10^{-12} \ {\rm GeV} \lsim m_\sigma \lsim 10^{-1} \ {\rm GeV}$, found previously in \hbox{Ref. \refcite{yeinzon}}, for the case where the curvaton oscillates for some time before decaying.

\section*{Acknowledgments}
The author is fully supported by COLCIENCIAS (COLOMBIA), and partially supported by COLFUTURO (COLOMBIA), UNIVERSITIES UK (UK), and the Department of Physics of Lancaster University (UK). Special acknowledgments are given to David H. Lyth who proposed this line of research and made comments on the manuscript.

\end{document}